\documentclass[aps,pre,twocolumn,showpacs,groupedaddress]{revtex4}
\usepackage{graphicx,psfrag,color}
\usepackage{bm}

\newcommand{\Tu}{T_1^{\mathrm {eq}}}
\newcommand{\Td}{T_2^{\mathrm {eq}}}
\newcommand{\bphi}{\bar \varphi}

\newcommand{\pq}[1]{\left[{#1}\right]}
\newcommand{\pg}[1]{\left\{{#1}\right\}}
\newcommand{\derpart}[2]{\frac{\partial #1}{\partial #2}}

\usepackage{comment}
\excludecomment{mycomments}

\begin{document}

\title{Out-of-equilibrium clock model at the verge of criticality}

\author{Marc \surname{Su\~n\'e}} 

\affiliation{Department of Physics and Astronomy, University of Aarhus\\ Ny 
Munkegade, Building 1520, DK--8000 Aarhus C, Denmark}
\author{Alberto Imparato}
\affiliation{Department of Physics and Astronomy, University of Aarhus\\ Ny 
Munkegade, Building 1520, DK--8000 Aarhus C, Denmark}

\date{\today}

\begin{abstract}
We consider an out-of-equilibrium lattice model consisting of 2D discrete rotators, in contact with heat reservoirs at different temperatures. The equilibrium counterpart of such model, the clock-model, exhibits three phases; a low-temperature ordered phase, a quasi-liquid phase, and a high-temperature disordered phase, with two corresponding phase transitions.
In the out-of-equilibrium model the simultaneous breaking of spatial symmetry and thermal equilibrium give rise to directed rotation of the spin variables.
In this regime the system behaves as a thermal machine converting heat currents into motion.
In order to quantify the susceptibility of the machine to the thermodynamic force driving it out-of-equilibrium, we introduce and study a dynamical response function. We show that the optimal operational regime for such a thermal machine occurs when the out-of-equilibrium disturbance is applied around the critical temperature at the boundary between the first two phases, namely where the system is mostly susceptible to external thermodynamic forces and exhibits a sharper transition.
We thus argue that critical fluctuations in a system of interacting motors can be exploited to enhance the machine overall dynamic and thermodynamic performances.

\end{abstract}
\maketitle

The critical behaviour of equilibrium systems is well understood and supported by innumerous experimental observations. As a system approaches a critical point larger fluctuations in the local order parameter take longer time to emerge and longer time to decay.
Also the response to a change in an external thermodynamic force can become significantly large. 
For example, in a gas at equilibrium with its liquid phase at the critical point a small change in the  pressure  produces a large density change, corresponding to a diverging compressibility.
Other well known examples of diverging responses are the specific heat and the susceptibility of a critical 3D Ising model.
Thus, criticality is characterized by wide fluctuations with diverging characteristic lengths,  susceptibilities, and relaxation times.

One may wonder what is the fate of the equilibrium critical fluctuations if a disturbance is applied, that drives the system out of equilibrium. The system must be certainly removed from criticality, but at least for small non-equilibrium perturbation one may expect that fluctuations in the system persist and are still quite large.
An even more interesting question is what happens to the diverging response of a critical system when such a non-equilibrium disturbance is applied.
Is the response of a system close to its critical points enhanced with respect to the case where the system is non critical?

This question is particularly relevant in the context of stochastic and quantum thermodynamics, where several groups have addressed the issue of performance optimization in microscopic machines operating as engines and motors \cite{Verley2014,Polettini15,Polettini17,SuneImparato19}. In particular, a few publications have shown that collective effects can enhance the efficiency of interacting microscopic machines.
In models of interacting molecular motors the efficiency at maximum power of the many-motor system can be larger than in the single motor case \cite{Golubeva2012a,Golubeva2013,Golubeva2014}. Models of interacting work-to-work transducers with all-to-all interaction exhibit a similar behaviour \cite{Imparato15,Herpich18,Herpich19}.
Furthermore, in ref.~\cite{Campisi2016} the authors showed that the Carnot efficiency can be approached at finite output power in a system of microscopic interacting Otto engines when the working substances operate at the verge of a second order phase transition and the critical exponents satisfy some given constraints. 


In this paper we investigate the role of criticality in the performance of thermal machines driven by temperature gradients: in particular we show that the out-of-equilibrium 
dynamic response of a system is maximal in the case in which the driving perturbation is applied 
at the verge of one of its critical points, i.e. when the system is most {\it susceptible} to equilibrium thermodynamic forces.
We also demonstrate that phase transitions with different strengths are characterized by different dynamical responses, with the smoother equilibrium phase transition exhibiting the feebler out-of-equilibrium response.

Specifically we consider a model of interacting autonomous thermal motors, that converts steady state heat currents into mechanical motion.
 Autonomous motors are systems which continuously and cyclically  convert  one form of energy into another: such machines are characterized by the absence of one or more  external agents that  change their Hamiltonian or temperature.
As such they are different from reciprocating engines that perform thermodynamic cycles, such as the Carnot, the Stirling and the Otto cycle, which have been widely investigated in the context of stochastic \cite{Seifert2012} and quantum \cite{Kosloff13} thermodynamics.

Our working tool will be the 2D clock model, consisting of $L\times L$ discrete rotors on a discrete 2D lattice, interacting through the Hamiltonian
\begin{equation}
H(\pg{\theta_i})=-\frac k 2 \sum_{\langle{i,j\rangle}}\cos\pq{\theta_i-\theta_j+\varphi},
\label{H:def}
\end{equation} 
with $\theta_i=2 \pi n_i/N_s$, $n_i=0,\dots N_s-1$, and with nearest neighbor interactions. We subdivide the lattice into two sublattices, each in contact with a heat bath at a different temperature $T_\pm=T_m\pm \Delta T$.
Specifically we take a 2D square-lattice with periodic boundary conditions and $L$ even, so that each ``+"  (``-") rotator has 4 nearest neighbour  ``-"  (``+") rotators \cite{suppmat}.
The model eq.~(\ref{H:def}) is chiral since we take $\varphi>0$ when the spin $i$ belongs to the ``-'' (cold) sublattice, and $j$ belongs to the "+" (hot) one.

This model has the advantage that i) a single dimer consisting of only two spins interacting with the Hamiltonian (\ref{H:def}) ($i=1,\, j=2)$ has been shown to work as an autonomous motor: when $\varphi\neq n \pi/N_s$ and when in contact with two heat baths at different temperatures the spins rotate with a given average frequency, thus converting heat currents into motion and  eventually into work \cite{Fogedby17,Hovhannisyan_2019}; and ii) its non-chiral equilibrium counterpart ($\varphi=0$, $\Delta T=0$) exhibits a rich phase diagram with two second-order phase transitions~\cite{Lapilli06}.
\begin{mycomments}
As we go into the details of the non-chiral equilibrium clock model later, here we just provide a brief description.
\end{mycomments}

\begin{mycomments}
{\it only one of the two critical points is ``good". CL about the second transition. The singularity at the KT critical temperature is a very weak essential one  (check). For example the specific heat exhibits an unobservable essential singularity, followed by a nonuniniversal peak associated with the entropy liberated by the unbinding of vortex pairs in the transition from the quasi-liquid to the disordered phase \cite{Chaikin} }
\end{mycomments}


In the following we use  Monte Carlo (MC) simulations with Metropolis algorithm to first investigate the equilibrium properties of the model~(\ref{H:def}) ($\Delta T=0$) and then its out-of-equilibrium dynamic properties, in particular the spin rotation rate for $\Delta T\neq 0$, and the heat rates exchanged by the system with the two reservoirs $\dot Q_\pm$.
We define the angular velocity per spin as 
$\omega=(\omega_++\omega_-)/2$, where $\omega_\pm$ is the angular velocity per spin in the hot (cold) sublattice.
 The heat currents $\dot Q_\pm$  along a single MC trajectory can be evaluated by sampling the total energy exchanged by the system with each of the two reservoirs along that trajectory.
\begin{mycomments}
{\it our model is different from \cite{Herpich18} and follow-up, as we have no all-to-all interaction. Genuine non-mean field phase transition. It is not a work-to-work transducer but a heat to work machine}
\end{mycomments}

\begin{figure}[h]
\center
\psfrag{ }[ct][ct][1.]{ }
\includegraphics[width=8cm]{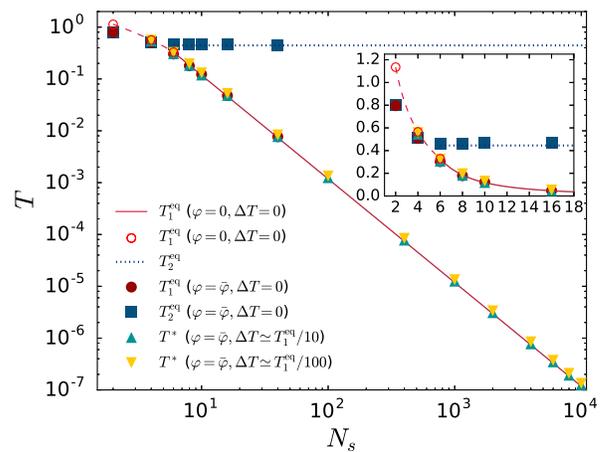}
\caption{Circles, squares, lines: equilibrium transition temperatures for the non-chiral ($\varphi=0$) and the chiral clock model ($\varphi\neq 0$) as given by eq.~(\ref{H:def}). Lines: the expressions for $\Tu$ (full) and $\Td$ (dotted) are taken from ref.~\cite{Lapilli06}. Open circles:  $\Tu$ for $N_s\le 6$, $\varphi=0, \, \Delta T=0$ ~\cite{Lapilli06}, the  dashed line is a guide to the eye for the results with $N_s\le 6$.
Full circles:  numerical results for the first transition temperature $\Tu$ as a function of $N_s$ ($N_s\le40$)  for the equilibrium model with $\varphi=\bphi$.
Squares: numerical results for the second transition temperature $\Td$ as a function of $N_s$, for the equilibrium model with $\varphi=\bphi$.  Triangles: temperature $T^*$ that maximizes the dynamic susceptibility $\chi_\omega$ in the out-of-equilibrium clock model with $\varphi=\bphi$, see text and figure~\ref{fig3}.
Inset: Zoom of the small $N_s$ region.
}
\label{fig1}
\end{figure}

We first study the equilibrium behaviour of the model (\ref{H:def}), and we benchmark our results with those discussed by {\it Lapilli et al.} in \cite{Lapilli06}, where the equilibrium properties of  the non--chiral clock model ($\varphi=0$) were extensively studied, and the model phase diagram was established by using MC simulations.
In particular, as discussed by the authors, the model (\ref{H:def}) with $4<N_s<\infty$ and $\varphi=0$ exhibits three phases: a low-temperature ordered and a high-temperature disordered phase, as in the Ising model,
along with a quasiliquid intermediate phase.
As $N_s\to \infty$ the ordered phase is suppressed, and the model becomes equivalent to the continuous $XY$ model. In the following we set the energy scale by taking $k=1$, and we notice that, because of the additional $1/2$ prefactor in our model (\ref{H:def}), the temperature scale in the present paper is halved with respect to \cite{Lapilli06}. We label $\Tu$ the equilibrium transition temperature between the low-temperature ordered and the quasi-liquid phase, and $\Td$ the equilibrium transition temperature between the quasi-liquid and the disordered phase. By using MC simulations, Lapilli {\it et al.} \cite{Lapilli06} provided an expression for the transition temperature $\Tu$ as a function of the number of states for $N_s\ge 6$.
They also reported an extended universality, which implies that the transition $\Td$ must be of the Berezinskii--Kosterlitz--Thouless (BKT) class, thus indistinguishable from $N_s=\infty$, for $N_s\geq 8$.
Such equilibrium temperatures are reported in fig.~\ref{fig1} as lines, together with the results from our MC simulations.

Similarly, we run equilibrium  MC simulations ($\Delta T=0$) for the chiral model $\varphi=\bphi $, where we fix $\bphi=\pi /(2 N_s)$, and evaluate the equilibrium transition temperatures $\Tu$ and $\Td$, for different values of $N_s$.  This is done  through finite-size scaling analysis of the order parameter $\langle m\rangle\equiv\langle|\sum_{l} \exp({i\theta_l})|\rangle/{L^2}$ (the sum runs over all the rotators in the system),  see \cite{suppmat, Lapilli06} for further details. 
The results for the two temperatures are also shown in fig.~\ref{fig1} and the agreement with the findings of \cite{Lapilli06} are excellent for $N_s\ge 8$.
This can be understood by noticing that the chosen phase $\varphi=\bphi$ decreases 
with $N_s$, so the equilibrium properties of the two models become equivalent in the limit of large $N_s$.

By using finite-size scaling analysis, we find that the critical exponent for the order parameter at the transition temperature $\Tu$ is a decreasing function of the number of states, thus signaling a sharper transition at $\Tu$ as $N_s$ increases, see Supplemental material~\cite{suppmat}.


 We now turn our attention to the model out-of-equilibrium properties.
As for the case of two isolated spins interacting with the Hamiltonian (\ref{H:def}) discussed in \cite{Hovhannisyan_2019}, also for the present model with $L^2$ spins there is a broken rotational symmetry when  $\varphi\neq n \pi/N_s$, see \cite{suppmat}.

\begin{figure}[h]
\center
\psfrag{ }[ct][ct][1.]{ }
\includegraphics[width=8cm]{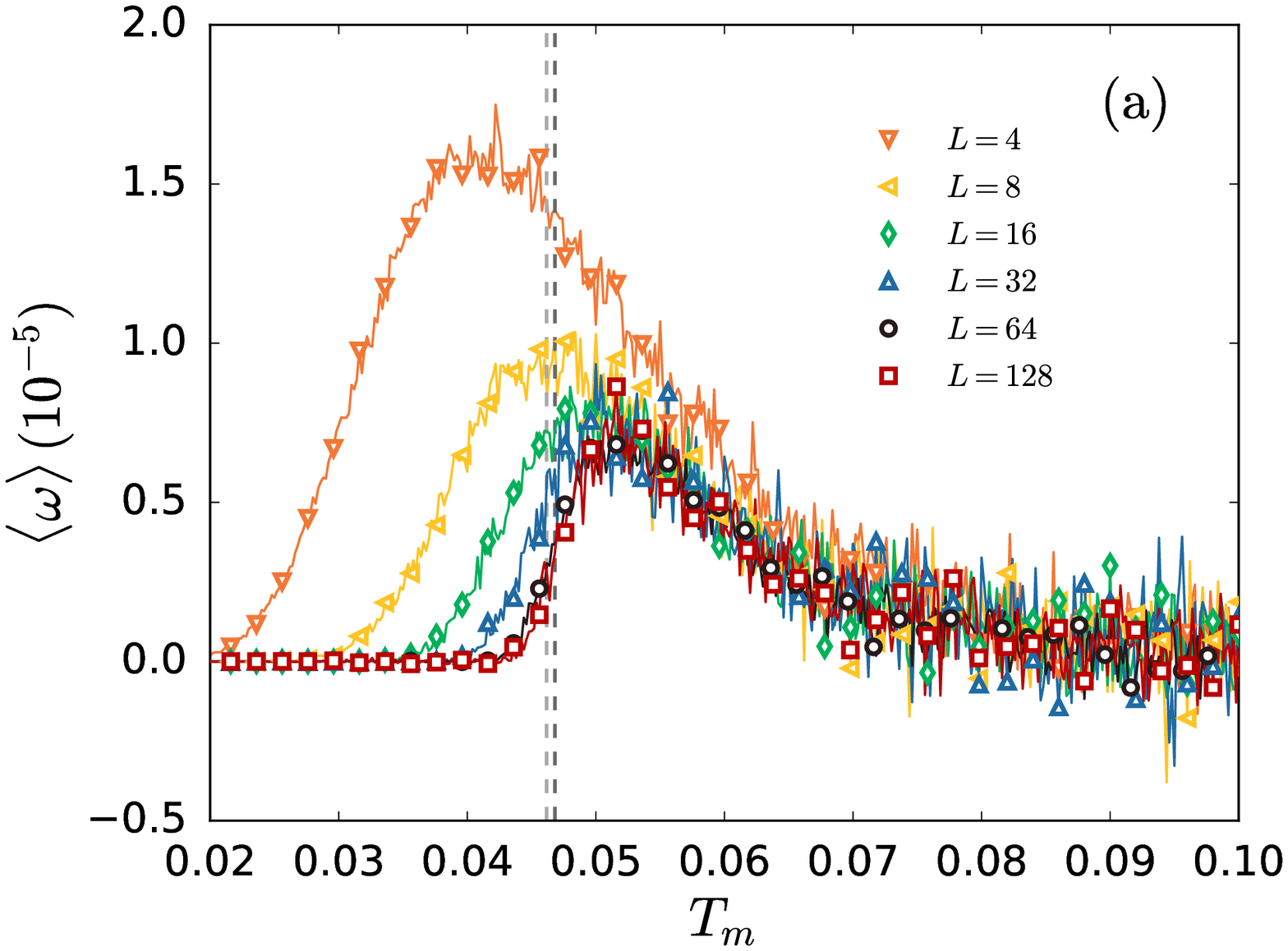}
\includegraphics[width=8cm]{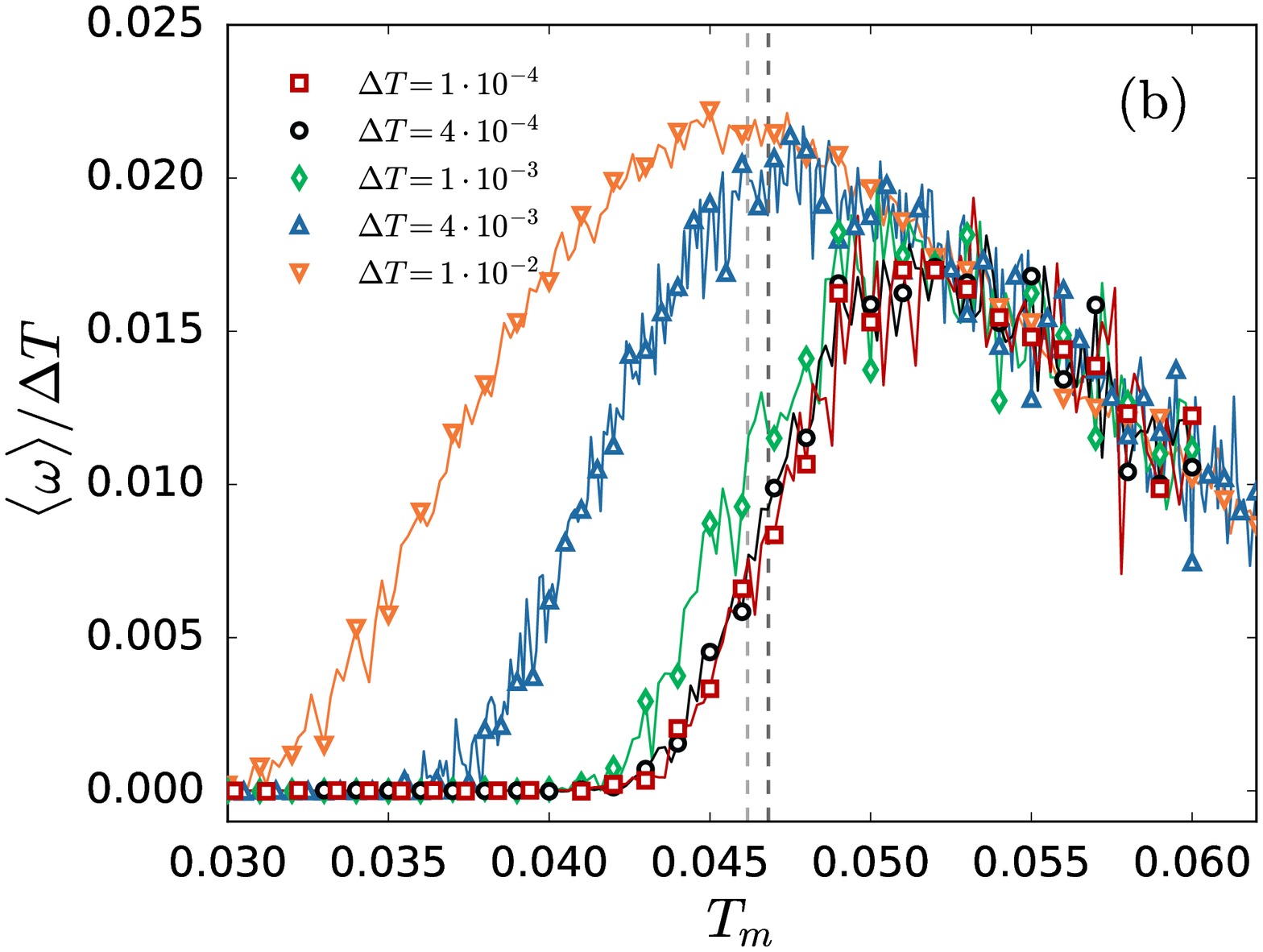}
\caption{(a) Average angular velocity per spin as a function of the mean lattice temperature $T_m$, for $N_s=16$ and $\Delta T=4\cdot 10^{-4}$.
(b) Ratio of the average angular velocity to the temperature gradient as a function of the mean lattice temperature $T_m$, for $N_s=16$ and $L=64$.
The vertical dashed lines correspond to the equilibrium transition temperature $\Tu$ for the clock model ($\varphi=0$, light gray)~\cite{Lapilli06}, and the chiral clock model ($\varphi\neq 0$, dark gray), see fig.~\ref{fig1}.
In both panels we considered trajectories consisting of 1000 $L^2$ MC steps. In (a) the results are obtained by averaging over ($10^6$, $2\cdot 10^5$, $5\cdot 10^4$, $8000$, $8000$, $1000$) independent trajectories for the different system sizes ($L=$4, 8, 16, 32, 64, 128). The angular velocity is expressed in units of 1/(number of MC steps per spin). In (b) the results are obtained by averaging over  ($10^4,8000,200,100,100$) trajectories for increasing $\Delta T$.}
\label{fig2}
\end{figure}

Thus, as in the 2 spin model \cite{Fogedby17, Hovhannisyan_2019}, when the broken rotational symmetry is combined with a constant temperature gradient, the system is driven into an out-of-equilibrium steady state, characterized by non-vanishing rotation rate per spin $\omega$.
The average angular velocity of the out-of-equilibrium system with $T_\pm=T_m \pm \Delta T$ is plotted in fig.~\ref{fig2} as a function of $T_m$ for $N_s=16$ and for different values of $L$ and $\Delta T$. We find that at large $L$ the curve $\omega(T_m)$ becomes independent of $L$, see fig.~\ref{fig2}-(a).

In order to investigate the role of the critical fluctuations in a non-equilibrium system we introduce a {\it dynamic response} to a non-equilibrium perturbation, akin to the susceptibility in equilibrium systems. In equilibrium statistical mechanics the susceptibility of an order parameter $\phi$ with respect to its conjugate thermodynamic force $f_\phi$ reads $\chi_\phi=\partial \phi/\partial f_\phi|_{f_\phi=0}$. 
Similarly we introduce the susceptibility 
\begin{equation}
\chi_\omega=\left. \derpart \omega {\Delta T}\right|_{\Delta T=0},\label{chi:def}
\end{equation} 
i.e., the response of the angular velocity with respect to the thermodynamic force that generates it. 
In the following we argue that this  is the relevant quantity to characterize the out-of-equilibrium response of a system close to its critical points.

Given that $\omega=0$ when $\Delta T=0$ we can thus evaluate the response function (\ref{chi:def}) by numerical derivation $\chi_\omega\simeq\omega/\Delta T$. 
Unless otherwise stated, in the following we take $\Delta T\simeq \Tu/100$, for any $N_s$. Indeed when $\Delta T$ is as small as $\Tu/100$ the curves $\omega(T_m)/\Delta T$ collapse into a single curve, see fig.~\ref{fig2}-(b).

\begin{figure}[h]
\center
\psfrag{ }[ct][ct][1.]{ }
\includegraphics[width=8cm]{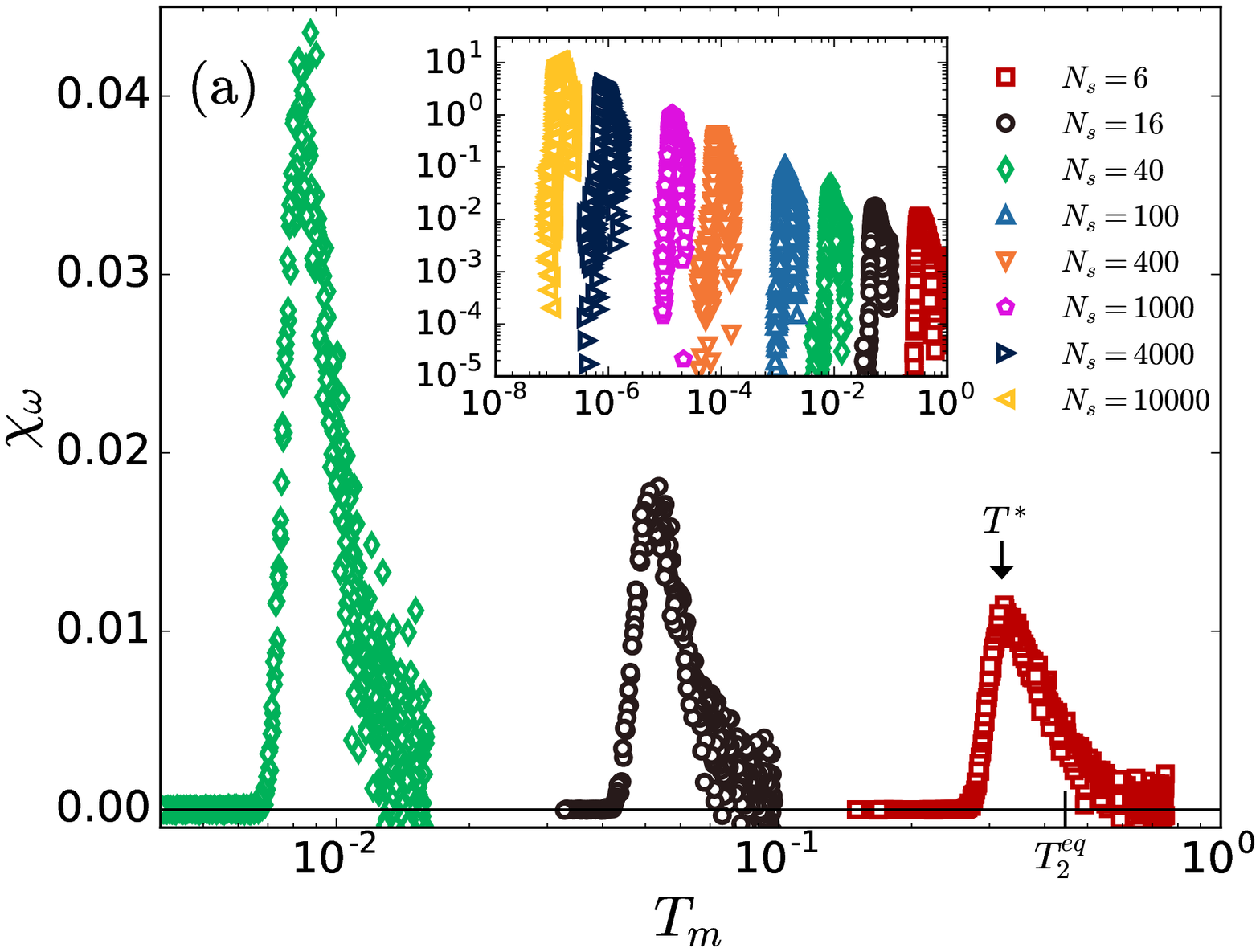}
\includegraphics[width=8cm]{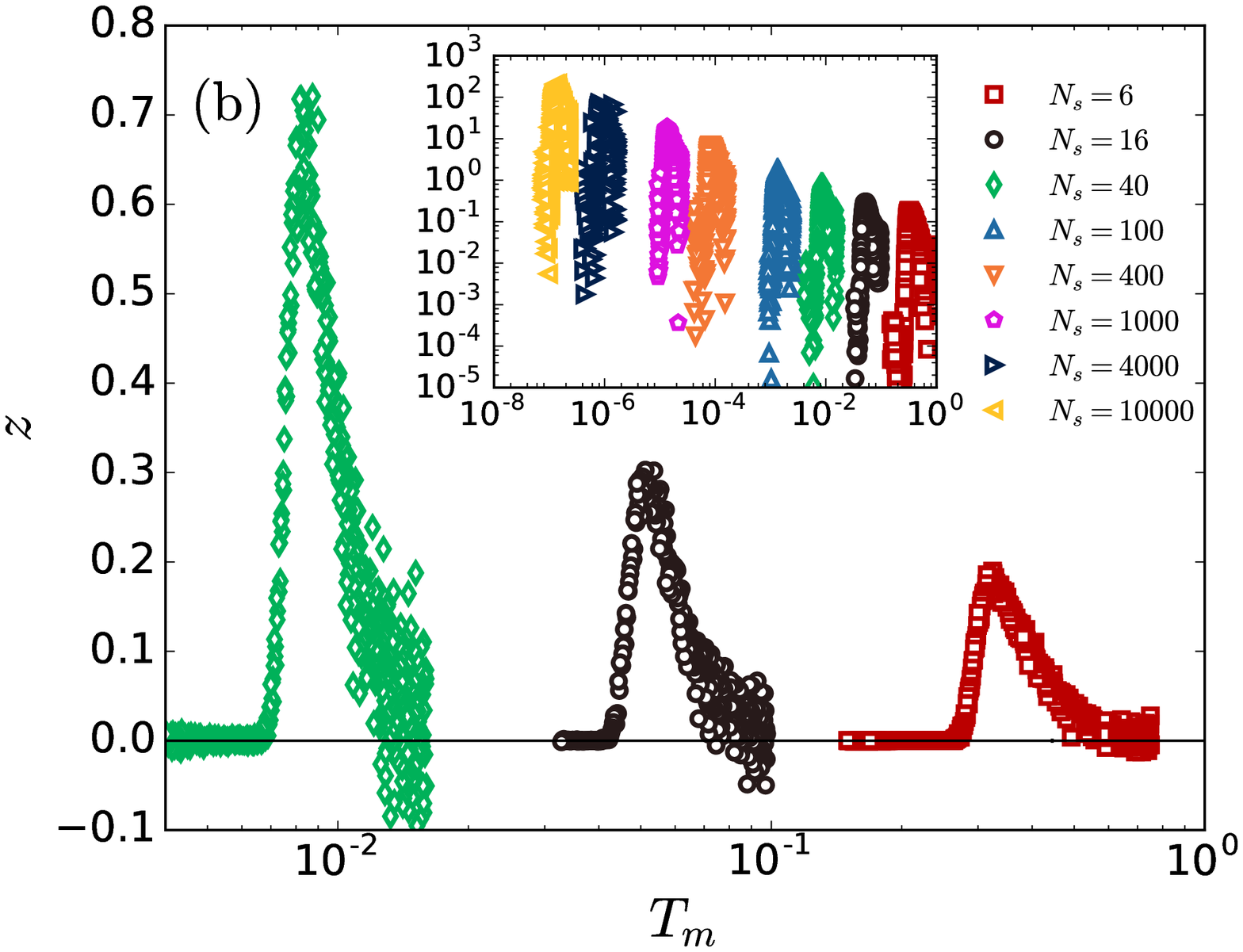}
\caption{(a) Out-of-equilibrium dynamical susceptibility $\chi_\omega$ as a function of the mean lattice temperature $T_m$, for different number of states $N_s$, $L=64$ and $\Delta T\simeq \Tu/100$.
We consider 8000 independent trajectory, each consisting of 1000 $L^2$ MC steps, after the system has reached its steady state.
For the system with $N_s=6$ we indicate the temperature $T^*$ for which  $\chi_\omega$ is maximum. The BKT temperature $\Td$, indicated in panel (a), is independent of the number of states $N_s$, see fig.~\ref{fig1}. Inset: $\chi_\omega$ as a function of $T_m$ for larger values of $N_s$.
(b) Ratio of the average angular velocity to the heat current $z=\langle  \omega  \rangle/\langle \dot Q_+ \rangle$, as a function of the mean temperature. Inset: $z$ as a function of $T_m$ for larger values of $N_s$.}
\label{fig3}
\end{figure}

The results for $\chi_\omega$ as a function of $T_m$ and for different values of $N_s$ are shown in fig.~\ref{fig3}-(a). We find that each curve $\chi_\omega(T_m,N_s)$ exhibits a peak at a temperature that depends on $N_s$. We indicate such a temperature $T^*$: it decreases as $N_s$ increases.
We refer to the value of the dynamic response function at its maximum as $\chi_\omega^*=\chi_\omega(T^*,N_s)$.

For any number of states $N_s$, we then compare the temperature $T^*$, as obtained for two values of $\Delta T$,  with the model equilibrium properties. In particular we compare the peak temperature with the two transition temperatures $\Tu$ and $\Td$. Interestingly, the temperature $T^*$ (triangles in fig.~\ref{fig1}) where $\chi_\omega(T_m,N_s)$ has its maximum $\chi_\omega^*$ 
follows very closely the equilibrium transition temperature $\Tu$ for $N_s\ge 4$ (full circles in fig.~\ref{fig1}) with a relative difference $(T^*-\Tu)/\Tu\lesssim 0.15$. This in turn implies that the points $T^*$ with $N_s\ge 6$ lie near the curve given in \cite{Lapilli06} for the equilibrium temperature $\Tu$ of the non-chiral model (full line in fig.~\ref{fig1}).
Accordingly, we conclude that the response of the system to the out-of-equilibrium perturbation $\Delta T$  is maximal when the perturbation is applied on the system in the proximity of one of its critical points: namely the transition point between the ordered and the quasiliquid phase.
A more precise analysis on the correspondence between $T^*$ and $\Tu$ can be found in~\cite{suppmat}.



Unlike $\Tu$, the temperature $\Td$ does not give rise to any nonvanishing response, for any value of $N_s$.
This might be seen as a consequence of the fact that the critical behaviour of the $XY$ model at the BKT transition exhibits very weak singularities \cite{Kosterlitz73,Chaikin}. For example the specific heat at $T_{\mathrm{ BKT}}$ displays a singularity which is essentially unobservable \cite{Chaikin}.
 
 In order to quantify the performance of the system, and its capability of transforming heat into rotational motion, we introduce the figure of merit $z= \langle \omega \rangle / \langle \dot Q_+ \rangle$, where $\langle \dot Q_+ \rangle$ is the steady state heat current entering the system from the hot reservoir.
Interestingly, such figure of merit follows the same behaviour of $\omega$ and $\chi_\omega$, that is, it peaks around $T^*$.
This result confirms that the optimal operational regime is achieved at $T_m\simeq\Tu$, see fig.~\ref{fig3}-(b).
This behaviour can be understood by noticing that while the angular velocity exhibits a somewhat abrupt peak around $T^*$ (figs.~\ref{fig2} and \ref{fig3}-(a)), the  heat current $\langle \dot Q_+ \rangle$ presents a smoother increase and decrease in that temperature range (data shown in \cite{suppmat}). Thus their ratio $z$ exhibits the same peaking behaviour as $\omega$ and $\chi_\omega$.

\begin{figure}[h]
\center
\psfrag{ }[ct][ct][1.]{ }
\includegraphics[width=8cm]{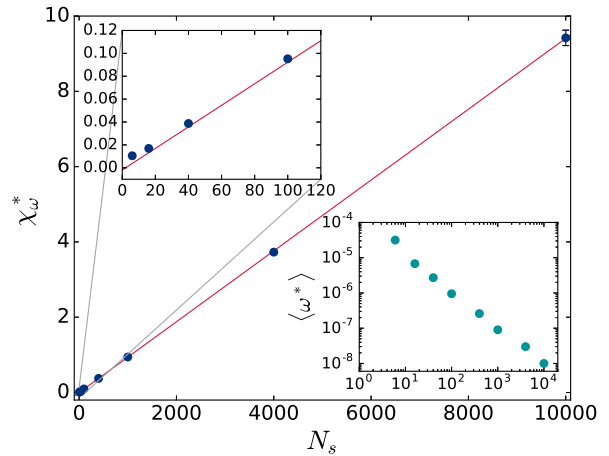}
\caption{Points: maximum value of the dynamical susceptibility $\chi_\omega^*$ as a function of the number of states $N_s$, from the data in fig.~\ref{fig3}-(a), as obtained with MC simulations for different number of states $N_s$, $L=64$, and $\Delta T\simeq\Tu/100$.
The full line is a linear fit to the data, with slope $b=(941.0\pm 1.6)\cdot 10^{-6}$.
Upper inset: zoom of the small $N_s$ region. Lower inset: average angular velocity $\langle\omega^*\rangle$ at $T^*$ as a function of the number of states. }
\label{fig4}
\end{figure}

Inspection of fig.~\ref{fig3} suggests that the height of the peak in both the dynamic susceptibility and the figure of merit is an increasing function of the number of states $N_s$. Thus as next point we investigate the dependency of $\chi_\omega^*$ on $N_s$; the results are plotted in fig.~\ref{fig4}. Such a quantity  increases linearly with the number of states, although with a slow rate. However, it is worth noticing that the velocity goes to zero as $N_s\to\infty$, as the rotational symmetry is restored (see \cite{suppmat}), as can be seen in the lower inset of fig.~\ref{fig4}. Therefore the continuous model $N_s\to\infty$ exhibits a diverging $\chi^*_\omega$, while $T^*\to 0$. 

Interestingly, the increase of $\chi_\omega^*$ is accompanied by the decrease of the order parameter critical exponent for the corresponding equilibrium system at $\Tu$, see ~\cite{suppmat}. We thus conclude that as the equilibrium transition becomes sharper, so does the corresponding out-of-equilibrium response.

To summarize, we have used the  2D chiral clock model to show that the optimal operational regime of a system of interacting autonomous motors occurs at the verge of a phase transition, in the limit of small driving thermodynamic forces.
While the out-of-equilibrium response around the ordered-quasiliquid phase transition is noticeably  sharp,   
the system response is vanishing in the other temperature ranges, in particular at the quasiliquid-disordered phase transition temperature.
The same behaviour is found if one considers a figure of merit akin to a miles per gallon equivalent: the motors' ``consumption"  is most favorable around the strong phase transition.
Interestingly, we find also that the response of the system increases, although slowly, with the number of states (directions) $N_s$ in the clock model.
This increase occurs while the corresponding phase transition for the equilibrium system becomes sharper.
Our results provide strong evidence for the protocol that one has to follow in order to find the optimal working regime in systems of microscopic interacting motors: one has to first identify the system critical points, if any, and then systematically perturb the system at the verge of criticality.

Experimentally, a 2D system of interacting molecular rotors deposited on an hexagonal lattice have been reported in \cite{Zhang2016}.
That artificial molecular system represents a possible setup for an experimental realization of the model we have studied here.
\begin{acknowledgments}
This work was supported by the Danish Council for Independent Research and the Villum Foundation.
The numerical  results presented in this work were obtained at the Centre for Scientific Computing, Aarhus http://phys.au.dk/forskning/cscaa.
\end{acknowledgments}

\appendix
\newpage
\section{Supplemental Material: Out-of-equilibrium clock model at the verge of criticality}

\subsection{Critical points for the  chiral clock model at equilibrium}
\label{sec:A1}
The Hamiltonian for the clock model in two dimensions eq.~(\ref{H:def}) with $\varphi=0$ has a global symmetry $Z_{N_s}$ ($H(\{\theta_i\})$ is invariant under the transformations $\theta_i\to\theta_i+2\pi n/N_s$).
Thus one introduces the order parameter
\begin{equation}
\langle m\rangle\equiv \frac{\langle M\rangle}{L^2}=\frac{\langle|\sum_{j=1}^{L^2} \exp({i\theta_j})|\rangle}{L^2},
\label{m:def}
\end{equation}
that is invariant under the same symmetry group as the Hamiltonian in the high-temperature disordered phase.
Besides this phase, the system exhibits two other phases for $N_s\ge 4$ (low-temperature ordered and quasi-liquid intermediate phase) and thus two critical points, as discussed in~\cite{Lapilli06}: an ordered to quasi-liquid second-order phase transition and a quasi-liquid to disordered phase transition, that turns out to be of the Berezinskii-Kosterlitz-Thouless class for $N_s\geq 8$.

\begin{figure}[h]
\center
\psfrag{ }[ct][ct][1.]{ }
\includegraphics[width=9cm]{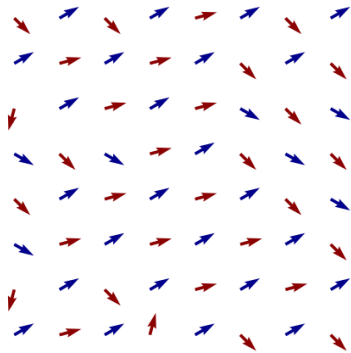}
\caption{Schematic representation of the 2D square-lattice model with $L=8$ and periodic boundary conditions. Each ``+" (``-") rotator in red (blue) has 4 nearest neighbour ``-" (``+") rotators in blue (red).}
\label{figA0}
\end{figure}

In the present paper we have considered the chiral model with $\varphi=\bar \varphi$. Similarly to the model with only two rotators  discussed  \cite{Karen18},  also for the present model with $L^2$ rotors  there is a broken rotational symmetry, in 
the sense that when $\varphi\neq n \pi/N_s$, one cannot find a discrete sublattice rotation  $l^+$ (or  $l^-$) such that $H(-\bm{\theta})=H(\{\theta^+_i+2 \pi l^+/N_s,\theta^-_j\})$, $\forall \bm \theta$ (or $H(-\bm{\theta})=H(\{\theta^+_i,\theta^-_j+2 \pi l^-/N_s\})$), where $\bm{\theta}=\{\bm\theta^+,\, \bm\theta^-\}$ is an arbitrary configuration of the system,  and  $\bm\theta^+,\, \bm\theta^-$ are the configurations of the two sublattices, see fig.~\ref{figA0}.
This symmetry plays a fundamental role in the out-of-equilibrium model described in the main text.
In the limit of $N_s\to \infty$ the model becomes the continuous $XY$ model and the energy eq.~(\ref{H:def}) exhibits an $O_2$ symmetry, the group of rotations in a two-dimensional plane, and thus the model does not break the spatial symmetry discussed above. The consequence is that the angular velocity for the out-of-equlibrium model vanishes as $N_s\to\infty$ as discussed in the main text.

\begin{figure}[h]
\center
\psfrag{ }[ct][ct][1.]{ }
\includegraphics[width=8cm]{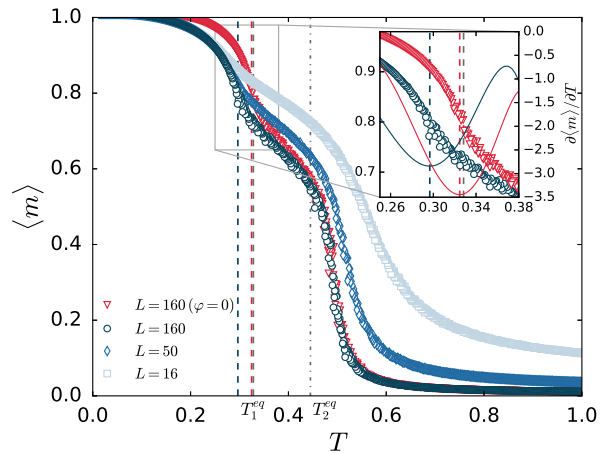}
\caption{Equilibrium order parameter for the standard clock model ($\varphi = 0$) and the chiral clock model ($\varphi=\bphi$) as given by eq.~(\ref{H:def}) for some values of $L$, and with $N_s=6$.
The colored vertical dashed lines depict the low-temperature transition point as given by the minimum of the order parameter's first derivative.
The gray vertical dashed and dashed-dotted lines correspond respectively to the equilibrium transition temperatures $\Tu$,$\Td$ of the non-chiral ($\varphi=0$) clock model obtained by {\it Lapilli et al.}~\cite{Lapilli06}.
Inset: Temperature derivative (full lines) of the equilibrium order parameter around the first transition temperature for $L=160$ for both $\varphi=0$ and $\varphi\bphi$.
We consider 100 independent trajectories each consisting of 1000 $L^2$ MC steps following an equilibration run of 1000 $L^2$ MC steps.}
\label{figA1}
\end{figure}

 Through equilibrium MC simulations of the model we have evaluated the order parameter (\ref{m:def}) as a function of the temperature. Such a quantity is shown in fig.~\ref{figA1} for $N_s=6$. Inspection of this figure indicates that the chiral model also exhibits two phase transitions. The order parameter for the standard clock model ($\varphi=0$) is included for comparison.

The order parameter curves display two abrupt drops as the temperature is increased, from the low-temperature ordered phase ($\langle m\rangle\approx 1$) to the high-temperature disordered phase ($\langle m\rangle\approx 0$).
In order to evaluate the corresponding transition temperatures we follow the approach used in ~\cite{Lapilli06}.
To obtain  the low-temperature transition $T_1^{eq}$,  we evaluate the   minimum of the temperature derivative of $m$ for different system sizes $L$, $T_1^{eq}(L)$, see inset in  fig.~\ref{figA1}.
We then use finite-size scaling analysis~\cite{PlischkeBergersen06} to obtain $T_1^{eq}=T_1^{eq}(L\to \infty)$.
%
%
%
For the high-temperature transition  $T_2^{eq}$ we exploit the Binder's fourth order cumulant~\cite{Binder02} of the order parameter,
\begin{equation}
U_L=1-\frac{\langle m^4\rangle}{3\langle m^2\rangle^2}.
\label{UL:def}
\end{equation} 
For $T> T_2^{eq}$ and assuming $L\gg \xi$, where $\xi$ is the correlation length of the order parameter fluctuations, $U_L\propto L^{-d}$ ($d$ is the dimension of the system).
On the other hand, $U_L\to U_{\infty}=2/3$ when $T< T_2^{eq}$ and $L\gg\xi$; and $U_L$ varies weakly with $T$ and $L$ for $L\ll\xi$.
Thus, the cumulant~(\ref{UL:def}) is scale invariant at the transition temperature, and therefore  the transition temperature is given by the intersection between ratios of cumulants for different system sizes~\cite{Binder02}.

The critical temperatures as computed employing the aforementioned approaches for the equilibrium chiral clock model are depicted in fig.~\ref{fig1}.
The second equilibrium transition temperature is the same for both the chiral version ($\varphi\neq0$) and the standard clock model ($\varphi=0$), as it can be anticipated by the collapse of the corresponding order parameters curves in fig.~\ref{figA1}.
However, the order parameter curves deviate as they approach the lower-temperature critical point, thus the chiral clock model exhibits the first transition at a lower temperature.
This deviation decreases with $N_s$ as the two models become equivalent in this limit. 

\begin{figure}[h]
\center
\psfrag{ }[ct][ct][1.]{ }
\includegraphics[width=8cm]{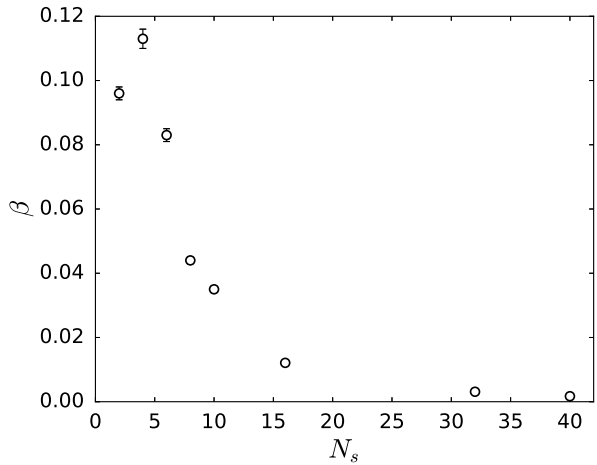}
\caption{Equilibrium chiral clock model ($\varphi=\bar \varphi$): critical exponent for the order parameter at the low temperature phase transition point $\Tu$, as a function of the number of states $N_s$.}
\label{figbeta}
\end{figure}

For different values $N_s$, and different sizes $L$, we  evaluate the Binder's fourth order cumulant for the energy \cite{Binder02}
\begin{equation}
V_L=1-\frac{\langle E^4\rangle}{3\langle E^2\rangle^2},
\label{VL:def}
\end{equation} 
around the first transition temperature $\Tu$. The latent heat is proportional to $\lim_{L\to \infty}(2/3- \min_T V_L)$ as discussed in \cite{Binder02}. We find that $\lim_{L\to \infty} \min_T V_L=2/3$, which thus signals a second-order phase transition.
We then proceed to evaluate the critical exponent $\beta$ for the order parameter $(m-m_c)\propto (T-\Tu)^\beta$, where $m_c$ is the value of $m$ at the transition point. We use finite-size scaling analysis of the order parameter for different sizes $L$~\cite{PlischkeBergersen06} to obtain the critical exponent $\beta$. The results are shown in fig.~\ref{figbeta}, where we plot the exponent $\beta$ as a function of $N_s$. We find that $\beta$ is a decreasing function of the number of states $N_s$, indicating that the transition becomes sharper as the number of states increases. This result has to be compared with the results for the dynamical susceptibility in the main text, whose maximum increases with $N_s$, see in particular fig.~\ref{fig4}.

\subsection{Correspondence between $T^*$ and $\Tu$}
\label{sec:A3}
\begin{figure}[h]
\center
\psfrag{ }[ct][ct][1.]{ }
\includegraphics[width=8cm]{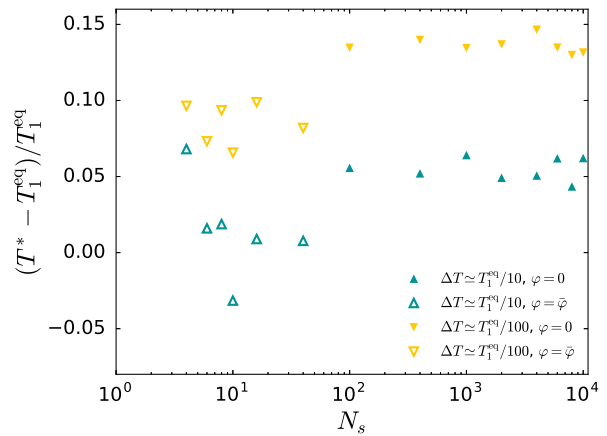}
\caption{Deviation of the temperature at the maximal susceptibility $T^*$ respect to the equilibrium clock model~\cite{Lapilli06} (filled triangles) and the equilibrium chiral clock model (open triangles) lower critical temperature $T_1^{\mathrm{eq}}$, for two different thermodynamic forces $\Delta T\simeq\Tu/10$ (green upward triangles) and $\Delta T\simeq\Tu/100$ (yellow downward triangles); $L=64$.}
\label{figA3}
\end{figure}

The temperature at which the susceptibility eq.~(\ref{chi:def}) is maximal $T^*$ has been shown to depend on $\Delta T$
, see fig.~\ref{fig2}-(b).
Its behavior when varying the number of states $N_s$ is depicted in fig.~\ref{fig1}, where two different temperature gradients are considered $\Delta T\simeq\Tu/10$ and $\Delta T\simeq\Tu/100$.
The maximum temperature $T^*$ follows closely the equilibrium clock model  transition temperature $T_1^{\mathrm{eq}}$ for both $\varphi=0,\, \bphi$), yet some discrepancy is already anticipated in fig.~\ref{fig1}.
In order to compare $T^*$ and $T_1^{\mathrm{eq}}$ a more convenient representation of the data in fig.~\ref{fig1} is to plot the relative deviation of the maximum susceptibility's temperature $T^*$ with respect to the equilibrium transition temperature $\Tu$
as a function of the number of states, see fig.~\ref{figA3}.

When decreasing the number of states, and hence when  the standard clock model begins to deviate from the chiral one, the maximal susceptibility temperature $T^*$ tends to approach the corresponding equilibrium temperature $\Tu$.
Yet  a deviation persists, so that $T^*$ is always above $T_1^{\mathrm{eq}}$ when the temperature gradient is small.
On the other hand, the difference between $T^*$ and $\Tu$ is larger for the smaller $\Delta T$ considered, in agreement with the results plotted in fig.~\ref{fig2}-(b).


\subsection{Heat current}
\label{sec:A2}
The heat currents $\dot Q_{\pm}$ along a single MC trajectory can be retrieved by exploiting the conservation of energy: the energy rate exchanged by each of the two sublattices with the corresponding heat reservoir must amount to the energy variation due to the sublattice's spin transitions, given by eq.~(\ref{H:def}).
Accordingly, the sum of the two heat currents 
should sum up to zero on average.

\begin{figure}[h]
\center
\psfrag{ }[ct][ct][1.]{ }
\includegraphics[width=8cm]{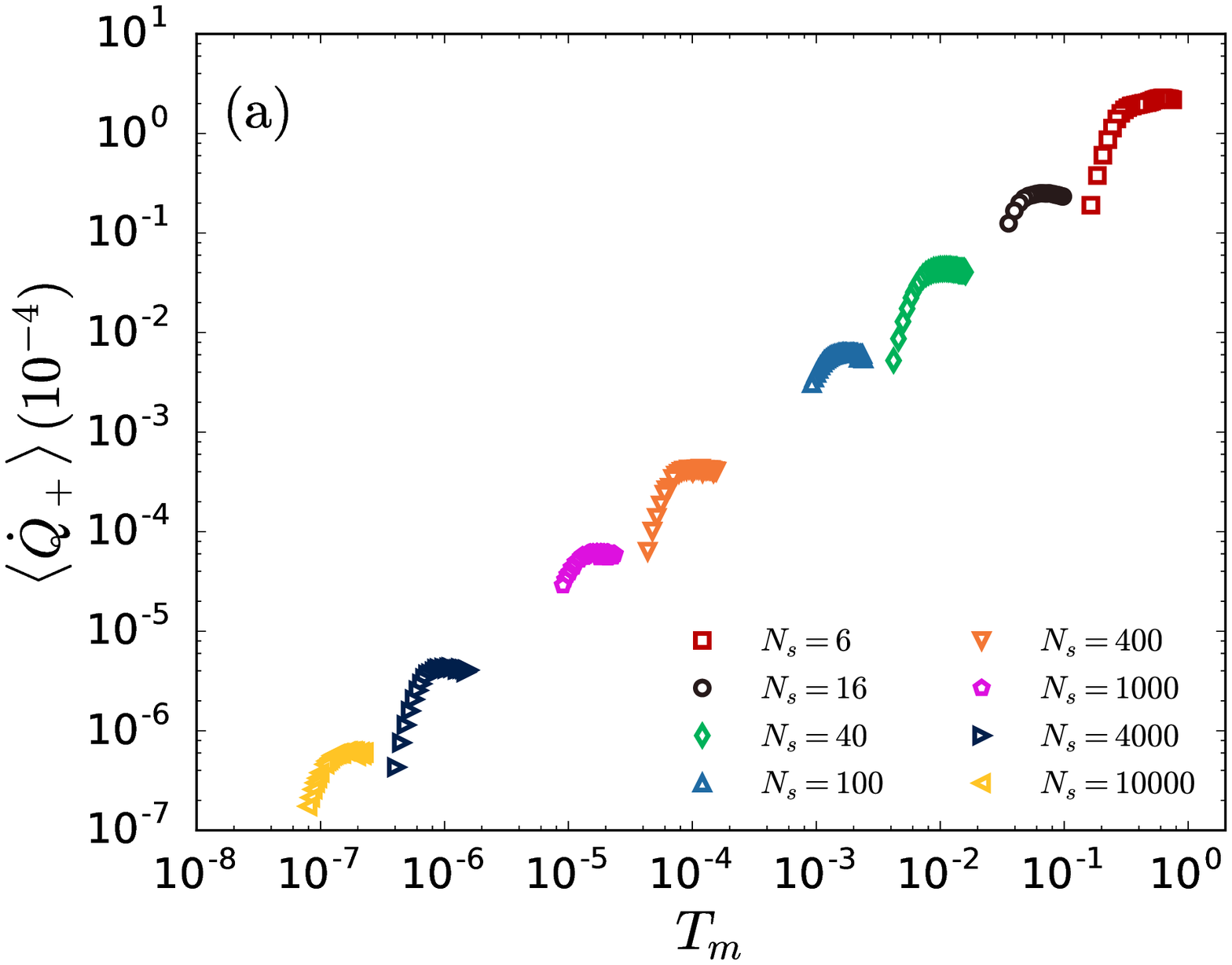}
\includegraphics[width=8cm]{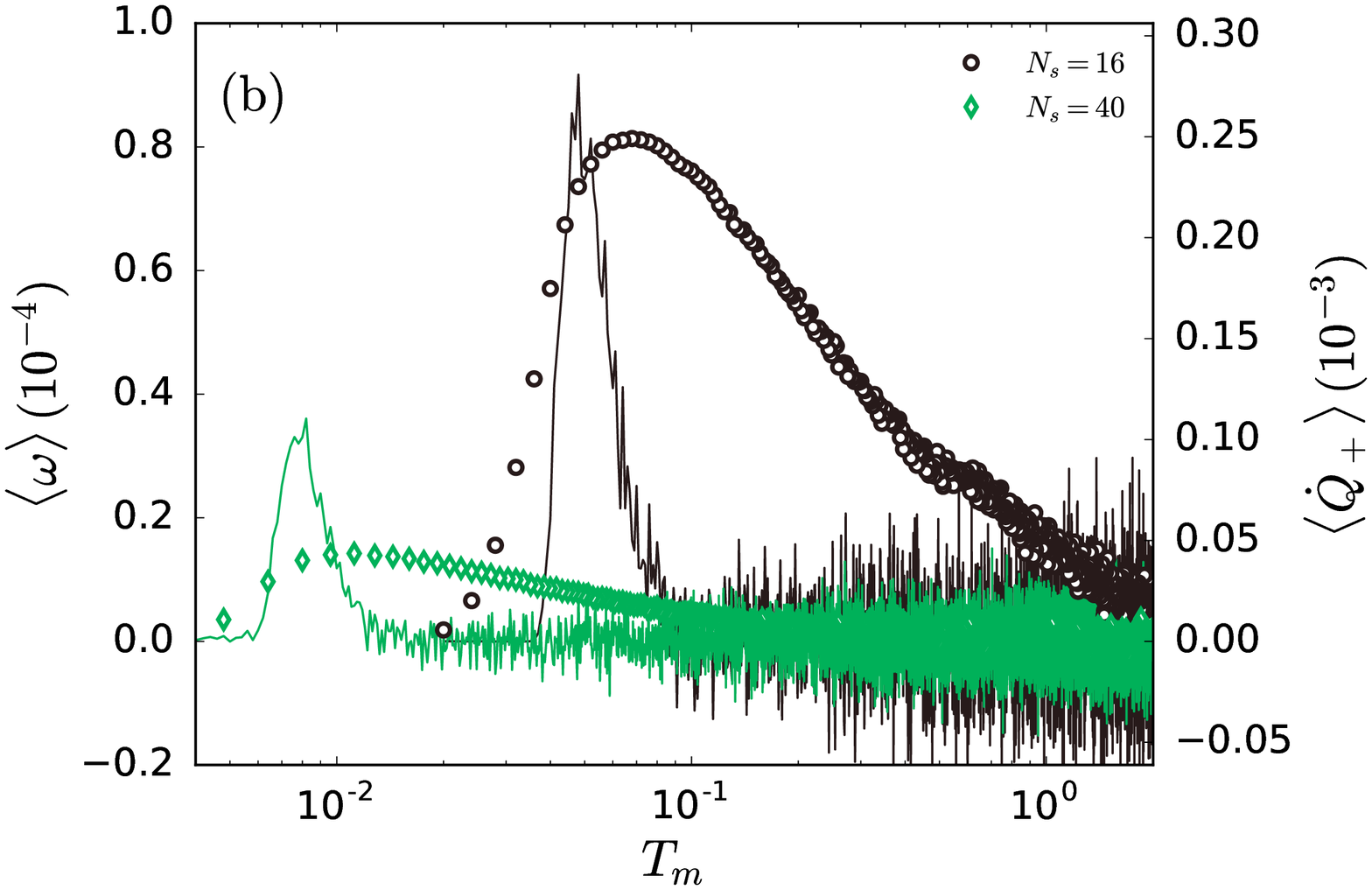}
\includegraphics[width=8cm]{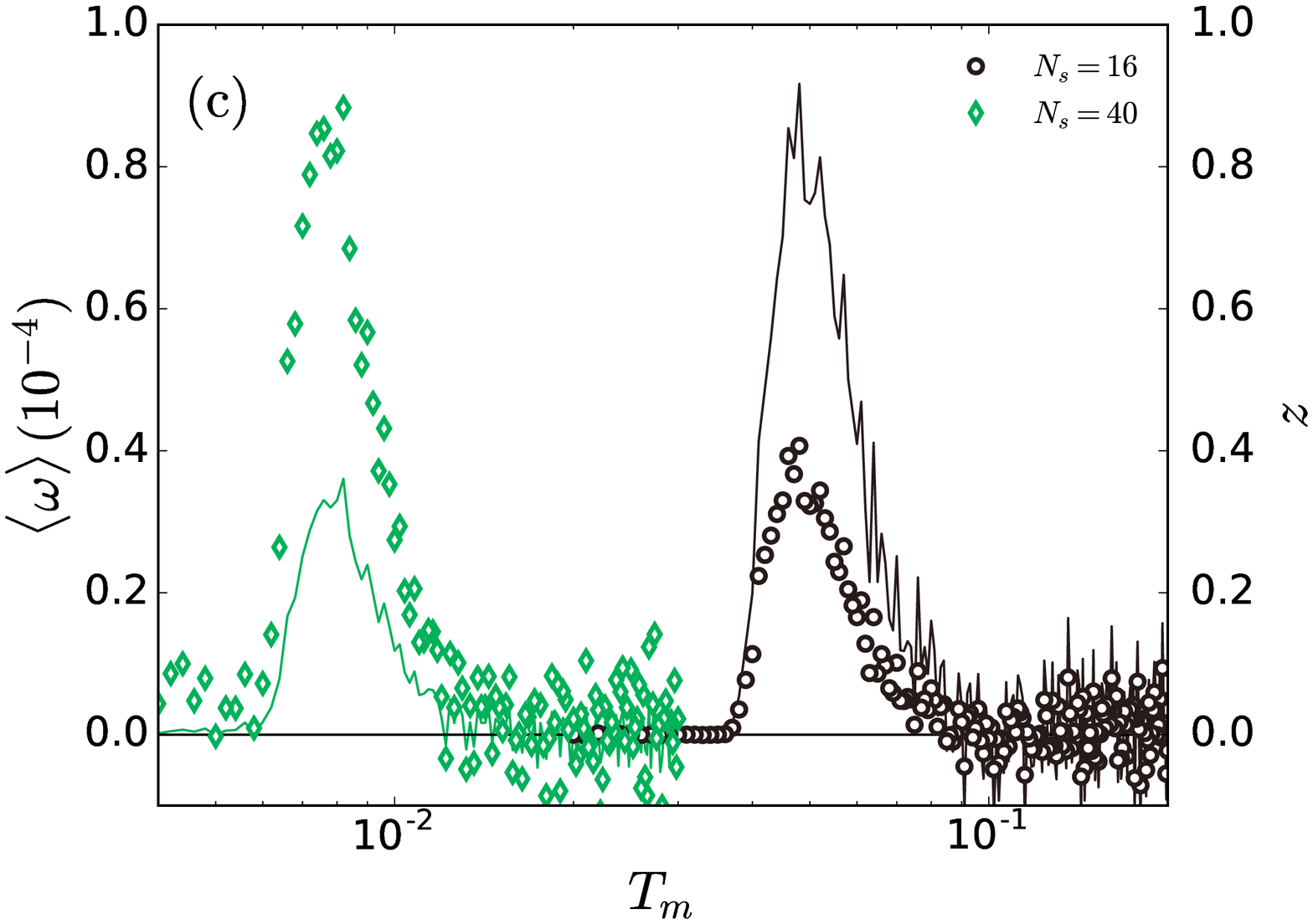}
\caption{(a) Average steady-state heat current flowing into the system from the hot reservoir $\langle\dot Q_+\rangle$ as a function of the mean lattice temperature $T_m$ for different number of states $N_s$, $L=64$ and $\Delta T\simeq \Tu/100$, corresponding to the simulations in fig.~\ref{fig3}.
(b) Average steady-state velocity $\langle\omega\rangle$ (lines, left y-axis) and heat current (symbols, right y-axis) flowing into the system from the hot reservoir as a function of the mean lattice temperature $T_m$ for $N_s=16,40$, $L=64$ and $\Delta T\simeq \Tu/10$.
(c) Comparison between the average steady-state velocity  (lines, left y-axis) and the corresponding figure of merit $z=\langle\omega\rangle/\langle\dot Q_+\rangle$ (symbols, right y-axis), same parameters as in panel (b).}
\label{figA2}
\end{figure}

The heat current as a function of $T_m$ for a given value of $N_s$  reaches a maximum which depends on $N_s$, similarly to what happens for the angular velocity, see fig.~\ref{figA2}-(a). Yet  a  close inspection shows that the curve reaches such a maximum at a  larger temperature than $T^*$, see fig.~\ref{figA2}-(b), where we compare the angular velocity $\langle\omega\rangle$ and the heat current $\langle\dot Q_+\rangle$ for two values of $N_s$.  The angular velocity exhibits a sudden rise/decrease around $T^*$. The heat current has a somewhat smoother rising and decreasing behaviour around the maximum. We see that the system still transports heat ($\langle\dot Q_+\rangle\neq 0$) in the range of $T_m$ where $\langle\omega\rangle$ vanishes due to the wandering of the rotators between their different positions, although this movement does not result in a net rotation.
The combination of the abrupt variation of $\langle\omega\rangle$ around $T^*$ together with the smoother behaviour displayed by $\langle\dot Q_+\rangle$ implies that the figure of merit $z$ (i.e. their ratio) follows the same behaviour of the angular velocity and peaks around $T^*$, see fig.~\ref{figA2}-(c).

\begin{figure}[h]
\center
\psfrag{ }[ct][ct][1.]{ }
\includegraphics[width=8cm]{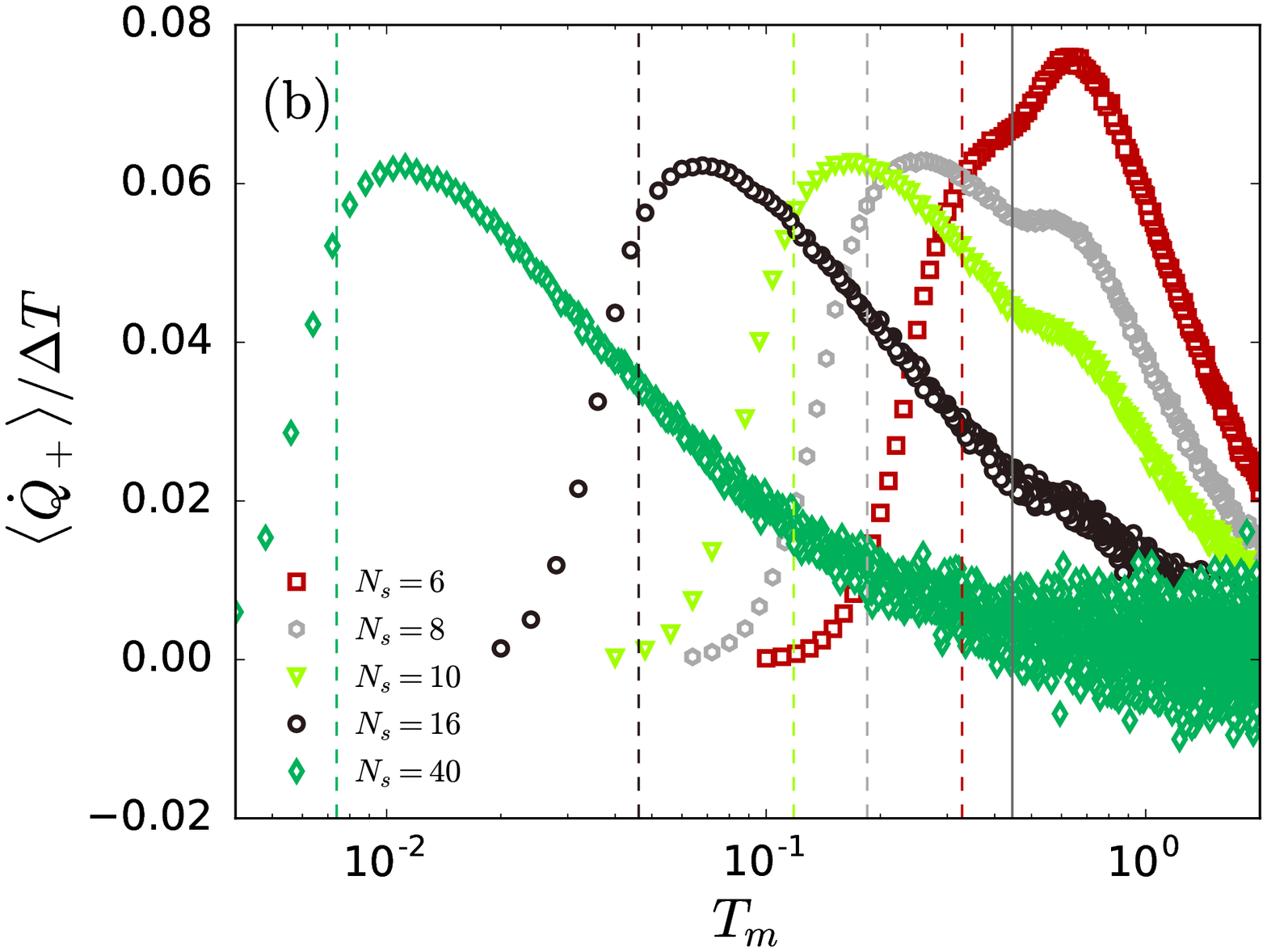}
\includegraphics[width=8cm]{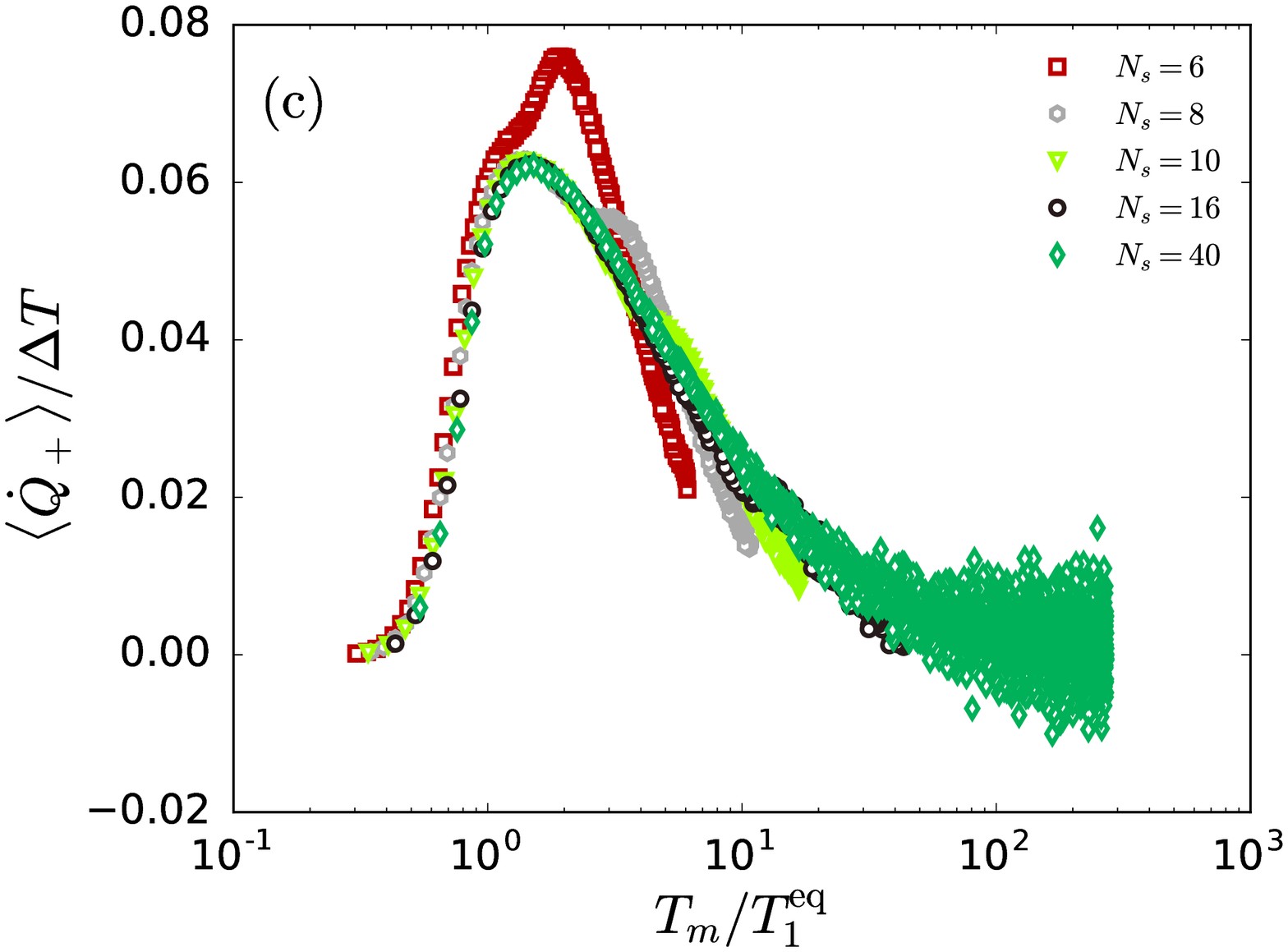}
\caption{Ratio of the average heat current flowing into the system from the heat reservoir to the out-of-equilibrium perturbation $\Delta T$ as a function of the mean temperature $T_m$ (a) and as function of $T_m/T_1^{\mathrm{eq}}$ (b) for different number of states $N_s$, $L=64$ and $\Delta T\simeq\Tu/10$. The coloured dashed lines and the full grey line in (a) label respectively the equilibrium transition temperatures $\Tu(N_s)$ and $\Td(N_s\geq 8)=T_{\mathrm{BKT}}$ for the standard clock model ($\varphi=0$) as obtained in~\cite{Lapilli06}.}
\label{figA5}
\end{figure}

The heat current's peak  at $T \gtrsim T^*$ is nonetheless not unique; a second maximum (possibly larger than the one discussed above) is found above $\Td$, see fig.~\ref{figA5}-(a), although only for small $N_s$.
The maxima displayed by the curves $\langle\dot Q_+\rangle/\Delta T$ close to the transition temperature $\Tu$ is constant with respect to the number of states $N_s$, see fig.~\ref{figA5}-(a), at variance with the mechanical current response function as defined in eq.~\ref{chi:def}, see also fig.~\ref{fig3}-(a) in the main text.
A rescaling of the {\it x}-axis thus leads to a collapse of the heat current curves around the transition temperature $T_1^{\mathrm{eq}}$, see fig.~\ref{figA5}-(b).
Both the mechanical and thermal currents are expected to be linear with respect to the thermodynamic force when $\Delta T$ is small, as it is in our case, according to the Onsager reciprocal relations.
However, we find a different scaling behavior of the two currents with respect to $N_s$, as suggested when comparing fig.~\ref{fig3}-(a) and fig.~\ref{figA5}-(a).

\end{document}